\documentclass[fleqn,usenatbib]{mnras}

\usepackage{newtxtext,newtxmath}
\usepackage{orcidlink}
\usepackage{verbatim}
 \usepackage[document]{ragged2e}

\usepackage[T1]{fontenc}
\usepackage{caption}

\DeclareRobustCommand{\VAN}[3]{#2}
\let\VANthebibliography\thebibliography
\def\thebibliography{\DeclareRobustCommand{\VAN}[3]{##3}\VANthebibliography}

\usepackage{amsmath}
\usepackage{xcolor}
\definecolor{my_color}{HTML}{3a18b1}

\definecolor{new_color}{HTML}{CF0000}
\definecolor{new_black}{HTML}{000000}
\usepackage{xfrac}

\newcommand\bedit[1]{\textcolor{new_black}{{#1}}}

\newcommand{\Kepler}{{\it Kepler}}

\newcommand{\TESS}{{\it TESS}}

\newcommand{\Gaia}{{\it Gaia}}
\newcommand{\rprs}{\ensuremath{r_p/r_\star}}

\newcommand{\Alopeke}{`$\!$Alopeke}

\newcommand{\be}{\begin{equation}}
\newcommand{\ee}{\end{equation}}

\newcommand{\mj}{M$_J$}

\newcommand{\ms}{\ensuremath{\rm m\,s^{-1}}}

\newcommand{\mearth}{M$_\oplus$}
\newcommand{\rearth}{R$_\oplus$}

\newcommand{\starone}{\bedit{K2-416}}
\newcommand{\staroneepic}{EPIC 246191231}
\newcommand{\distancestarone}{$120.77\pm0.58$}
\newcommand{\massstarone}{$0.548\pm0.012$}
\newcommand{\radiusstarone}{$0.554\pm0.017$}
\newcommand{\loggstarone}{$4.691\pm0.038$}

\newcommand{\efftempstarone}{$3746\pm72$}
\newcommand{\luminositystarone}{$0.0537\pm0.0013$}
\newcommand{\periodstarone}{10.1\ensuremath{^{+10.2}_{-3.5}}}

\newcommand{\transittimestarone}{2458371.8347\ensuremath{^{+0.0041}_{-0.0051}}}
\newcommand{\rprstarone}{0.0445\ensuremath{^{+0.0049}_{-0.0024}}}
\newcommand{\impactstarone}{0.52\ensuremath{^{+0.32}_{-0.35}}}
\newcommand{\transitdurationstarone}{2.63\ensuremath{^{+0.29}_{-0.38}}}

\newcommand{\arstarone}{27.1\ensuremath{^{+5.7}_{-9.5}}}
\newcommand{\aaustarone}{0.0690\ensuremath{^{+0.015}_{-0.024}}}
\newcommand{\incstarone}{88.9\ensuremath{^{+0.8}_{-1.6}}}
\newcommand{\rpstarone}{2.7\ensuremath{^{+0.3}_{-0.17}}}

\newcommand{\periodstaronesecondtransit}{13.1020\ensuremath{^{+0.0026}_{-0.0024}}}
\newcommand{\transittimestaronesecondtransit}{2458371.8384\ensuremath{^{+0.0018}_{-0.0021}}}
\newcommand{\rprstaronesecondtransit}{0.0440\ensuremath{^{+0.0029}_{-0.0018}}}
\newcommand{\impactstaronesecondtransit}{0.45\ensuremath{^{+0.29}_{-0.30}}}
\newcommand{\transitdurationstaronesecondtransit}{2.77\ensuremath{^{+0.13}_{-0.10}}}

\newcommand{\arstaronesecondtransit}{34.2\ensuremath{^{+3.8}_{-8.0}}}
\newcommand{\aaustaronesecondtransit}{0.088\ensuremath{^{+0.010}_{-0.021}}}
\newcommand{\incstaronesecondtransit}{89.25\ensuremath{^{+0.53}_{-0.86}}}
\newcommand{\rpstaronesecondtransit}{2.66\ensuremath{^{+0.19}_{-0.14}}}

\newcommand{\startwo}{\bedit{K2-417}}
\newcommand{\startwoepic}{EPIC 245978988}
\newcommand{\distancestartwo}{$94.30\pm0.30$}
\newcommand{\massstartwo}{$0.569\pm0.012$}
\newcommand{\radiusstartwo}{$0.5776\pm0.0040$}
\newcommand{\loggstartwo}{$4.670\pm0.014$}

\newcommand{\efftempstartwo}{$3861\pm77$}
\newcommand{\luminositystartwo}{$0.0628\pm0.0010$}
\newcommand{\periodstartwo}{8.3\ensuremath{^{+7.5}_{-3.0}}}

\newcommand{\transittimestartwo}{2458373.0288\ensuremath{^{+0.0030}_{-0.0059}}}
\newcommand{\rprstartwo}{0.0504\ensuremath{^{+0.0035}_{-0.0022}}}
\newcommand{\impactstartwo}{0.45\ensuremath{^{+0.30}_{-0.30}}}
\newcommand{\transitdurationstartwo}{2.45\ensuremath{^{+0.32}_{-0.24}}}

\newcommand{\arstartwo}{24.1\ensuremath{^{+4.5}_{-5.6}}}
\newcommand{\aaustartwo}{0.065\ensuremath{^{+0.012}_{-0.015}}}
\newcommand{\incstartwo}{89.0\ensuremath{^{+0.7}_{-1.2}}}
\newcommand{\rpstartwo}{3.18\ensuremath{^{+0.22}_{-0.14}}}

\newcommand{\periodstartwosecondtransit}{6.5350\ensuremath{^{+0.0014}_{-0.0014}}}
\newcommand{\transittimestartwosecondtransit}{2458373.0321\ensuremath{^{+0.0009}_{-0.0010}}}
\newcommand{\rprstartwosecondtransit}{0.0525\ensuremath{^{+0.0022}_{-0.0019}}}
\newcommand{\impactstartwosecondtransit}{0.41\ensuremath{^{+0.29}_{-0.28}}}
\newcommand{\transitdurationstartwosecondtransit}{2.07\ensuremath{^{+0.10}_{-0.08}}}

\newcommand{\arstartwosecondtransit}{23.4\ensuremath{^{+2.3}_{-5.1}}}
\newcommand{\aaustartwosecondtransit}{0.063\ensuremath{^{+0.006}_{-0.014}}}
\newcommand{\incstartwosecondtransit}{89.00\ensuremath{^{+0.71}_{-1.20}}}
\newcommand{\rpstartwosecondtransit}{3.31\ensuremath{^{+0.14}_{-0.12}}}

\newcommand{\starthree}{EPIC 246251988}
\newcommand{\distancestarthree}{$361.9\pm5.3$}
\newcommand{\massstarthree}{$1.099\pm0.038$}
\newcommand{\radiusstarthree}{$1.586\pm0.045$}
\newcommand{\loggstarthree}{$4.24\pm0.10$}

\newcommand{\efftempstarthree}{$5721\pm50$}
\newcommand{\luminositystarthree}{$2.43\pm0.16$}
\newcommand{\periodstarthree}{10.6\ensuremath{^{+11.5}_{-2.9}}}

\newcommand{\transittimestarthree}{2458375.6752\ensuremath{^{+0.0035}_{-0.0038}}}
\newcommand{\rprstarthree}{0.0219\ensuremath{^{+0.0023}_{-0.0014}}}
\newcommand{\impactstarthree}{0.53\ensuremath{^{+0.34}_{-0.36}}}
\newcommand{\transitdurationstarthree}{3.25\ensuremath{^{+0.310}_{-0.23}}}

\newcommand{\arstarthree}{21.8\ensuremath{^{+4.0}_{-9.2}}}
\newcommand{\aaustarthree}{0.16\ensuremath{^{+0.030}_{-0.068}}}
\newcommand{\incstarthree}{88.6\ensuremath{^{+1.0}_{-2.6}}}
\newcommand{\rpstarthree}{3.8\ensuremath{^{+0.42}_{-0.270}}}

\title[No Planets Left Behind]{\Kepler's Last Planet Discoveries: Two New Planets and One Single-Transit Candidate from K2 Campaign 19}

\author[Incha et al.]{
Elyse Incha,$^{1,2}$\orcidlink{0000-0001-5637-6144} \thanks{E-mail: eincha@wisc.edu}
Andrew Vanderburg,$^3$\orcidlink{0000-0001-7246-5438}
Tom Jacobs$^4$\orcidlink{0000-0003-3988-3245}, Daryll LaCourse$^5$\orcidlink{0000-0002-8527-2114}, Allyson Bieryla$^6$\orcidlink{0000-0001-6637-5401},\newauthor Emily Pass$^6$\orcidlink{0000-0002-1533-9029},
Steve B. Howell$^{7}$\orcidlink{0000-0002-2532-2853}, Perry Berlind$^{6}$, Michael Calkins$^{6}$\orcidlink{0000-0002-2830-5661}, Gilbert Esquerdo$^{6}$\orcidlink{0000-0002-9789-5474},
\newauthor David W. Latham$^{6}$\orcidlink{0000-0001-9911-7388}, 
Andrew W. Mann$^{8}$\orcidlink{0000-0003-3654-1602},  
\\
$^{1}$Department of Astronomy, University of Wisconsin-Madison, 475 N. Charter St., Madison, WI 53703, USA\\
$^{2}$UW-Madison Sophomore Research Fellow and Hilldale Research Fellow\\
$^{3}$Department of Physics and Kavli Institute for Astrophysics and Space Research, Massachusetts Institute of Technology, Cambridge, MA 02139, USA\\
$^{4}$Amateur Astronomer, 12812 SE 69th Place, Bellevue, WA 98006, USA\\
$^{5}$Amateur Astronomer, 7507 52nd Pl NE, Marysville, WA 98270, USA \\
$^{6}$Center for Astrophysics \textbar \ Harvard \& Smithsonian, 60 Garden Street, Cambridge, MA 02138, USA\\
$^{7}$NASA Ames Research Center, Moffett Field, CA 94035, USA\\
$^{8}$Department of Physics and Astronomy, The University of North Carolina at Chapel Hill, Chapel Hill, NC 27599, USA\\
}

\date{Accepted XXX. Received YYY; in original form ZZZ}

\pubyear{2023}

\begin{document}
\label{firstpage}
\pagerange{\pageref{firstpage}--\pageref{lastpage}}
\maketitle
\justify
\begin{abstract}
\justify 
The \Kepler\ space telescope was responsible for the discovery of over 2,700 confirmed exoplanets, more than half of the total number of exoplanets known today. These discoveries took place during both \Kepler's primary mission, when it spent 4 years staring at the same part of the sky, and its extended K2 mission, when a mechanical failure forced it to observe different parts of the sky along the ecliptic. At the very end of the mission, when \Kepler\ was exhausting the last of its fuel reserves, it collected a short set of observations known as K2 Campaign 19. So far, no planets have been discovered in this dataset because it only yielded about a week of high-quality data. Here, we report some of the last planet discoveries made by \Kepler\ in the Campaign 19 dataset. We conducted a visual search of the week of high-quality Campaign 19 data and identified three possible planet transits. Each planet candidate was originally identified with only one recorded transit, from which we were able to estimate the planets' radii and estimate the semimajor axes and orbital periods. Analysis of lower-quality data collected after low fuel pressure caused the telescope's pointing precision to suffer revealed additional transits for two of these candidates, allowing us to statistically validate them as genuine exoplanets. We also tentatively confirm the transits of one planet with TESS. These discoveries demonstrate \Kepler's exoplanet detection power, even when it was literally running on fumes.  

\end{abstract}




\section{Introduction}

\begin{figure*} 
   \centering
   \includegraphics[scale=.65]{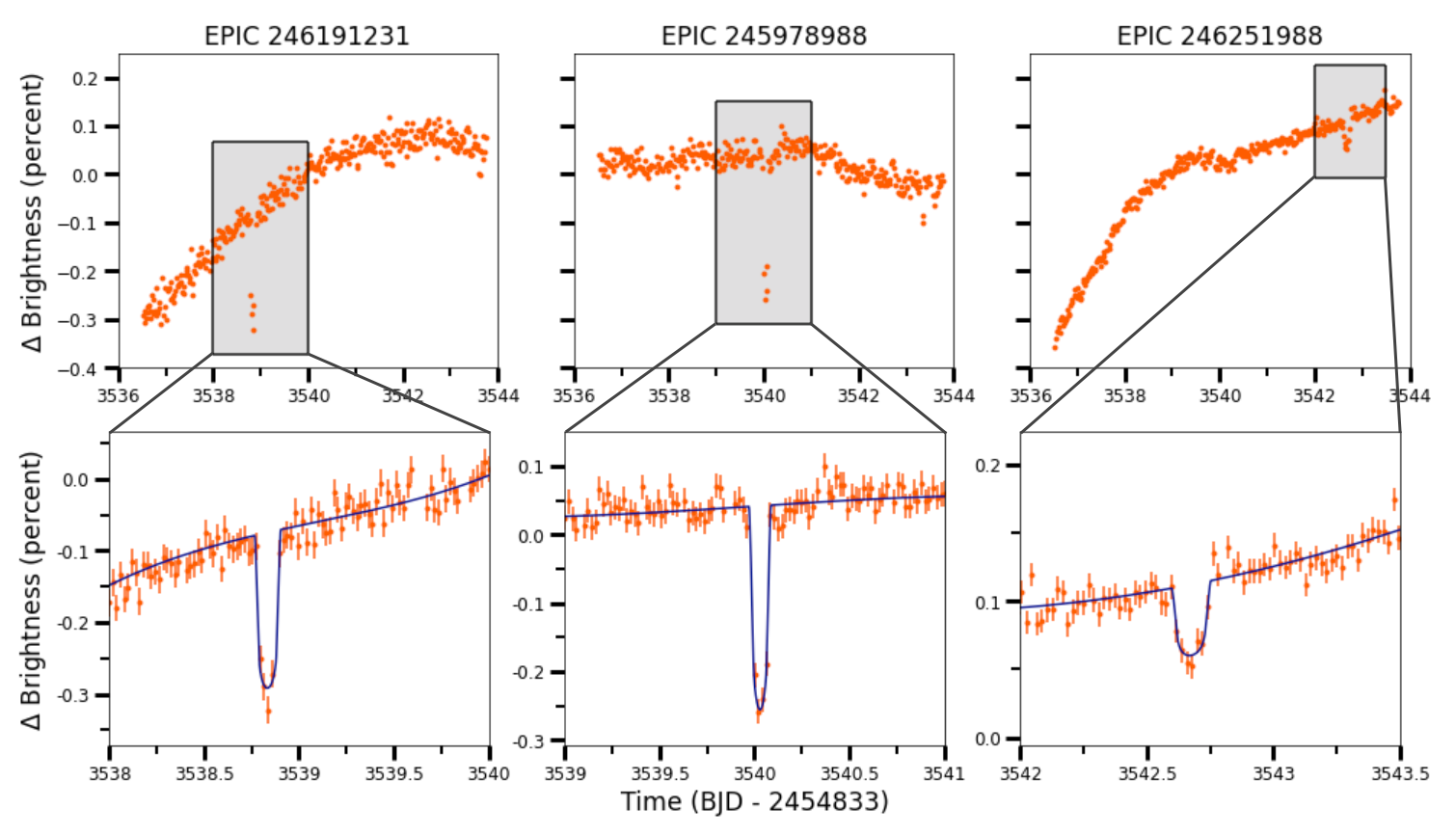} 
   \caption{Light curves of each of the three planet candidate systems. \textit{Top row:} The full 7.25-day high-quality discovery light curves of each of our systems are shown on the top, produced via the process outlined in Section \ref{lightcurve}. \textit{Bottom row:} Zoomed-in light curves near the three single transits, along with the best-fit transit model from Section \ref{transitmodel}. The best fit parameters are outlined in Table \ref{bigtable}. We detected additional transits of two of these three planet candidates, which are shown in Figure \ref{newtransits}. \bedit{We also note that there are two small dipping features in the light curve of \startwoepic\ that can be seen around t=3538 days and t=3543.5 days. We conclude that these are systematic because they are not robust to changing data reduction parameters.}}
   \label{lightcurves_bestfits}
\end{figure*}

With its 110 square degree field of view and 1 meter aperture, \Kepler\ was a powerful observatory designed with the primary goal of finding Earth analog exoplanets. \Kepler\ operated by taking images of thousands of pre-selected stars every 30 minutes, and even in its first few months of operations was able to produce the most precise light curves astronomers had ever seen \citep{keplerfirstsixweeks}. After its launch in 2009, \Kepler\ observed over 200,000 stars during its primary mission and had discovered thousands of exoplanet candidates by May of 2013 \citep{batalha2013}. These discoveries revolutionized our understanding of individual planetary systems \citep{lissauerkepler11, carter, morton2016}, planetary system architectures \citep{rowe, fabrycky}, and overall demographics \citep{fressin, dressing, gapinradii}.

\Kepler\ relied on its four reaction wheels -- small, gyroscope like devices -- to keep it precisely pointed towards its original observation field located in the constellation Cygnus. When its first wheel failed in July of 2012, the \Kepler\ team was still able to perform its science operations with the remaining three. When the second wheel failed in May 2013, however, there was no way to continue to orient the telescope stably towards its original field of view, so the primary \Kepler\ mission ended. Fortunately, this was not the end for \Kepler, as the team members from Ball Aerospace and NASA were able to come up with a new operation mode that enabled the telescope to point somewhat precisely at different parts of the sky (antisolar regions along the ecliptic) for a shorter period of time. This new concept became known as the K2 mission.

During K2, the \Kepler\ telescope used its two remaining reaction wheels to orient the center of its field of view and balanced at an unstable equilibrium against the Sun's radiation pressure using its thrusters to maintain working equilibrium \citep{K2plan}. Because \Kepler\ was only able to maintain this balance in one direction, which changed as the telescope moved in its orbit, the telescope observed different fields of view of throughout its orbit. The geometry of the spacecraft's balance point against solar radiation pressure restricted the observations to fields along the ecliptic plane. Sets of observations of a given field, called campaigns, were limited to a duration of about 80 days at a time.

The K2 mission operated between 2014 and 2018, completing almost 20 campaigns\footnote{The campaigns began with Campaign 0.} in which it discovered over a thousand planet candidates, hundreds of which have been confirmed or validated as genuine exoplanets \citep{vanderburg16, pope, santerne, crossfieldcatalog, kruse, zink, christiansen}. However, because K2 relied on thrusters to maintain its pointing (unlike during the primary mission, when pointing was fully maintained with reaction wheels), thruster fuel expenditure increased significantly and ultimately limited the mission's lifetime. K2 completed Campaigns 0-17 and over half of Campaign 18 before the mission operations team noticed a sudden decline in the fuel tank pressure gauge. To avoid losing the data collected up to that point, the team prematurely terminated Campaign 18. After successfully recovering the Campaign 18 data, the \Kepler\ team started Campaign 19, but the fuel reserves were so low that the campaign only lasted about a month and only yielded about 7.25 days of high-quality data. Shortly afterwards, the team decommissioned \Kepler\ and ended the mission. 

However, just because a set of \Kepler\ observations had a short duration did not mean they could not bear scientific fruit. Due to \Kepler's extreme photometric precision, even short observation periods were enough to produce cutting-edge scientific results. During the first observations of its original field, \Kepler\ detected the optical phase curve and secondary eclipse of HAT-P-7 b \citep{keplerhatp7} in just 10 days of data. After the failure of \Kepler's second reaction wheel, an 11-day engineering test at the beginning of the K2 mission to determine the viability of the new operating mode enabled white dwarf astroeseismology \citep{hermes2014} and led to the discovery of numerous eclipsing binaries \citep{conroy} and the K2 mission's first planet discovery \citep{HIP116454}. Even short glimpses of new targets with \Kepler's unprecedented photometric precision were enough to reveal new details about the planetary systems in its field of view. 

In this paper, we provide another example of the power of \Kepler\ by describing two planets and one planet candidate identified in the short-lived Campaign 19 dataset, collected during September 2018 immediately before the telescope completely ran out of fuel and was decommissioned. We searched the 7.25-day high-quality dataset for exoplanet transits, and identified promising candidates around three stars: \staroneepic, \startwoepic, and \starthree.  Due to the campaign's short duration, each planet candidate only transited once during the 7.25 days of good observations, but we were able to recover additional transits for two of the three planet candidates in data collected after the data quality deteriorated. Using these additional transits, we refine the parameters for \staroneepic\ b and \startwoepic\ b and validate them as genuine exoplanets called \starone\ b and \startwo\ b, respectively.  

\begin{figure*}
      \centering
      \includegraphics[scale = 0.50]{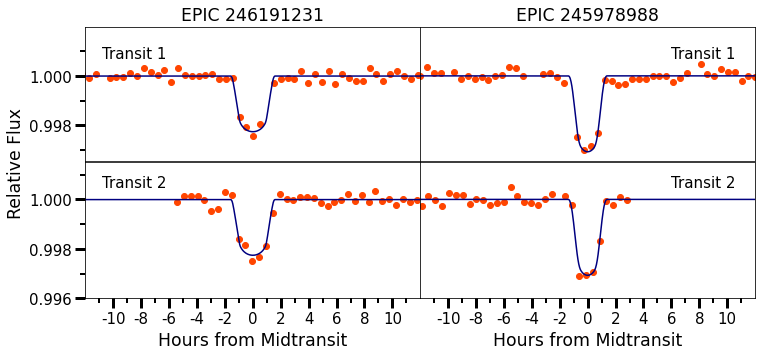}
      \caption{Light curves of additional transits detected in the \staroneepic\ and \startwoepic\ light curves. \textit{Top row:} The original single transits detected in the 7.25 days of high-quality data. \textit{Bottom row:} Additional transits found in data collected after high-quality data collection was no longer possible, as described in in Section \ref{lightcurve}.  The best-fit transit model from a combined fit to the two transits is shown overlaid in blue. The shapes, depths, and durations of the first and second transits in each light curve are highly consistent, as expected for subsequent transits of the same planet.}
      \label{newtransits}
   \end{figure*}

Our paper is organized as follows. First, we describe how we detected these new planet candidates in Section \ref{observations} and the data we collected and used in our investigation.  Section \ref{analysis} outlines the ways in which we modeled these transits (Section \ref{transitmodel}), determined their stellar and planetary parameters (Sections \ref{stellarparameters}, \ref{transitmodel}, and \ref{radialvelocitymodelling}). We ruled out other astrophysical phenomena which may prove the transits to be false positives and validate \starone\ b and \startwo\ b in Section \ref{validation} (using constraints determined in Sections \ref{archivalimaging},  \ref{speckleimaging}, and \ref{radialvelocitymodelling}). We discuss our results in Section \ref{discussion}, and conclude in Section \ref{summary}.

\section{Observations}\label{observations}

\subsection{\Kepler\ Light Curve}\label{lightcurve}


\staroneepic, \startwoepic, and \starthree\ were observed by \Kepler\ during Campaign 19 of the K2 mission from 2018 Aug 29 to 2018 Sep 26, along with more than 33,000 other targets. Data collection during Campaign 19 was challenging for the spacecraft due to the extremely low thruster fuel levels. As a result, \Kepler\ struggled to acquire its targets at the beginning of the campaign, and operated for the first 8.5 days of the campaign with the pointing off-center. Eventually, \Kepler\ was able to properly center its targets in the field of view, and observed normally for about 7.25 days. After those 7.25 days, the spacecraft's motion became increasingly erratic as the thrusters were only able to execute burns intermittently to correct the pointing drift. This behavior continued for about 11 days. During this time, certain parts of the light curve were still useable, but the pointing was too erratic \bedit{and the spacecraft's field of view moved too far off center} for general purpose pipelines to function well. At the end of these 11 days, mission team concluded that \Kepler\ had run out of sufficient fuel for precise photometric observations. At that time, observations ceased and all data recorded was downloaded to Earth. Of the 27 days of data recorded, only about 7.25 days were suitable for reduction with existing pipelines. \Kepler\ did not retain enough fuel for a 20th Campaign, making Campaign 19 the telescope's final set of observations.

Once downloaded, the Campaign 19 data were processed by the K2 mission pipeline and made publicly available via the K2 data archive on the Barbara A. Mikulski Archive for Space Telescopes (MAST). Our team accessed the Campaign 19 data from MAST and processed them into light curves following the process outlined in \citet{vanderburglightcurve2014}, focusing on the 7.25 days of precise photometry collected in the middle of the campaign. This process removed systematic errors from \Kepler's instability, leaving behind clean light curves. 

The light curves of all 33,000 stars were searched by members of our team by eye using a process outlined in \citet{tomanddaryll} and \citet{kristiansen}. By this method, we were able to identify single transits of planet candidates orbiting three stars: \staroneepic, \startwoepic, and \starthree. Once we identified these planet candidates, we re-processed the K2 light curves making several improvements to yield the best photometric precision and fewest biases due to the systematics correction process. The processing had two main differences from our standard processing. First, we used a circular moving aperture to extract the light curves of these three stars, to minimize the effects of the increasingly erratic telescope pointing \citep[as done by][]{curtis}. Second, we derived the systematics correction by simultaneously fitting a transit model along with the K2 roll systematics and a long-term trend for each star \citep{vanderburg16}. This method yielded the light curves shown in Figure \ref{lightcurves_bestfits}.

After identifying the three single-transit planet candidates, we attempted to reduce the data collected during the final 11 days of Campaign 19 after K2 lost the ability to point normally due to its diminishing fuel reserves. These data are challenging to analyze and interpret because the erratic behavior of the thrusters caused significantly larger and less regular pointing drifts -- and therefore light curve artifacts -- that caused our standard methods for systematics correction to fail. However, useful data from this time has been recovered on a star-by-star basis \citep{sha}.

We therefore produced light curves from the final 11 days of Campaign 19. We did not attempt to perform any type of systematics correction, instead opting to identify the times of thruster fires and visually inspect the light curves between these events. K2 light curves typically exhibit sharp jumps at the times of thruster firing events, with smooth curves in between, so we looked for transit-like dips in flux away from the times of thruster firing events. We saw no transit like signals in the light curve of \starthree, but in the light curves for \staroneepic\ and \startwoepic,  we identified transits with the same depth, duration, and shape as the ones detected earlier in the light curve (see Figure \ref{newtransits}). These transit-like features were unique in their respective un-corrected light curves away from the times of thruster firing events and were robust to different choices of photometric aperture. We confirmed that our candidate periods were likely to be the true orbital period of the planet, and not an integer multiple, by inspecting the data one half,  one-third, and two-thirds of the time between the two detected transits and finding no additional transits at those times. We therefore conclude that we detected additional transits for the candidates orbiting \staroneepic\ and \startwoepic\ and measured their precise orbital periods.


\begin{table}
	\begin{center}
	\caption{Radial velocity observations of \staroneepic, \startwoepic, and \starthree.}
\label{rvvalues}
	\begin{tabular}{lcc} 
		\hline
		Star (EPIC ID) & Time (BJD) & RV (m/s)\\
		\hline
246191231 & 2459472.872354 & $-9287.1 \pm 100.7$  \\
246191231 & 2459521.76389 & $-9239 \pm 60.1$\\
246191231 & 2459522.726005 & $-9230.3 \pm 50.0$ \\
245978988 & 2459471.834055 & $19394 \pm 64.9$ \\
245978988 & 2459518.69286 & $19313.1 \pm 59.7$ \\
245978988 & 2459521.726594 & $19516.2 \pm 60.3$ \\
245978988 & 2459522.762002 & $19476.6 \pm 52.0$ \\
246251988 & 2459157.749659 & $-19572.0 \pm 96.2$ \\
246251988 & 2459158.680957 & $-19429.4 \pm 45.8$ \\
246251988 & 2459164.686084 & $-19337.3 \pm 45.8$ \\
246251988 & 2459190.638456 & $-19386.5 \pm 43.6$ \\
246251988 & 2459195.580134 & $-19311.2 \pm 46.1$ \\
		\hline
	\end{tabular}
	\end{center}
	\justify{\textit{Notes: } \staroneepic, \startwoepic, and \starthree\ were all observed with the Tillinghast Reflector Echelle Spectrograph (TRES) on the 1.5 m telescope at Fred L. Whipple Observatory with a resolving power $\Delta \lambda / \lambda=44000$. \starthree\ was observed in November of 2020 while \staroneepic\ and \startwoepic\ were observed in November of 2021. Because these stars are all fairly dim, we were only able to obtain a few radial velocity measurements for each star, listed above. We fit model radial velocity curves to these data in Section \ref{radialvelocitymodelling}.}
\end{table}

\subsection{Archival Imaging} \label{archivalimaging}

After detecting the transits, we inspected archival imaging observations to identify or rule out the presence of background stars at the present-day locations of \staroneepic, \startwoepic, and \starthree. We did this by comparing historic observations taken decades ago with more recent observations of the same parts of the sky. Because the universe around us is constantly in motion, over decade time scales we can see the stars in our sky move in relation to each other.  This tactic is useful in ruling out the possibility that the transits we observed could have been false positives caused by background eclipsing binaries. 

In particular, we took advantage of the fact that all three of our stars were observed by the First Palomar Sky Survey (POSS I, \citealt{abell1955}) in the 1950s which utilized the 48-inch Oschin Schmidt telescope at Palomar Mountain in California. More recently, the stars were also observed by the Panoramic Survey Telescope and Rapid Response System (Pan-STARRS, \citealt{chambers}) in the 2010's which is operated by the Institute for Astronomy at the University of Hawaii. Images of each star from both surveys can be seen in Figure \ref{star_images}. We found that both \staroneepic\ and \startwoepic\ have high enough proper motion that their position in the sky has moved far enough in the past seven decades to reveal the sky behind their present day positions. We observed no stationary background objects behind \staroneepic\ or \startwoepic, ruling out background stars as a possible source of the transit signal. \starthree, however, is a and more distant G-type star and has a smaller proper motion. The star hasn't moved enough to determine whether it is hiding stationary background objects. We cannot rule out interference from background eclipsing binaries in the case of \starthree.

\begin{figure*}
    \centering
    \includegraphics[scale=0.35]{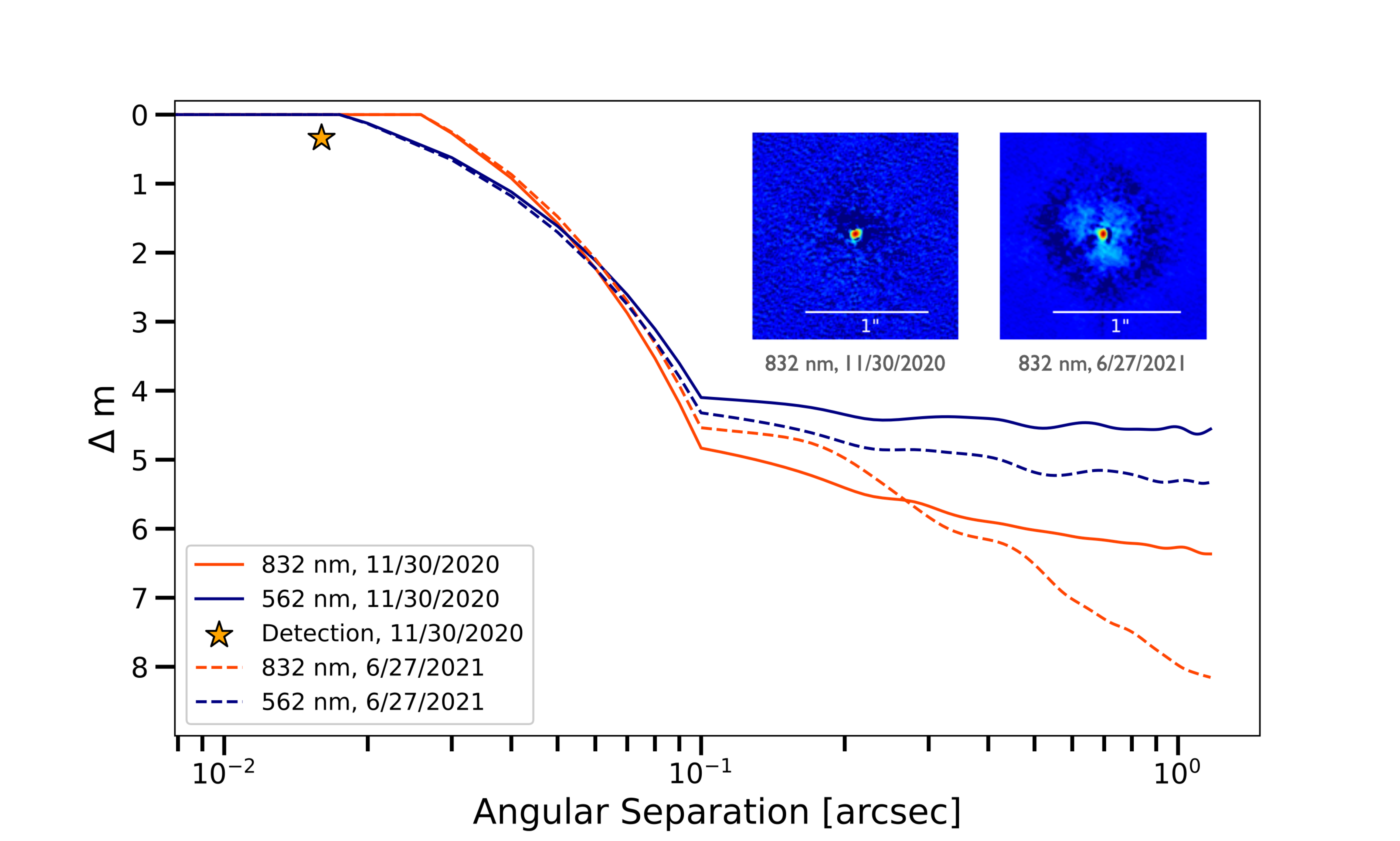}
    \caption{\bedit{Contrast curves obtained from speckle observations of \starthree. The solid lines represent data observed using the Zorro instrument at Gemini-South on October November 30th, 2020. In this observation, a possible companion detection was made at $\Delta m=0.34$ and an angular separation of 0.016 arcsec in the 832 nm filter (shown as an orange star in the plot). Because this possible companion was detected at Gemini Observatory's sensitivity limit (inside the nominal contrast curve) and only detected in one filter, another observation was taken on June 27th, 2021 with Gemini North (plotted above with dashed lines). These observations revealed no detection. The orange lines above represent the contrast curves from the 832 nm filter and the blue lines represent contrast curves from the 562 nm filter. The speckle images reconstructed from the 832 filter observations can be seen in the top right corner of the figure.}}
    \label{specklefigurestarthree}
\end{figure*}

\begin{figure*}
   \centering
   \includegraphics[scale=.4]{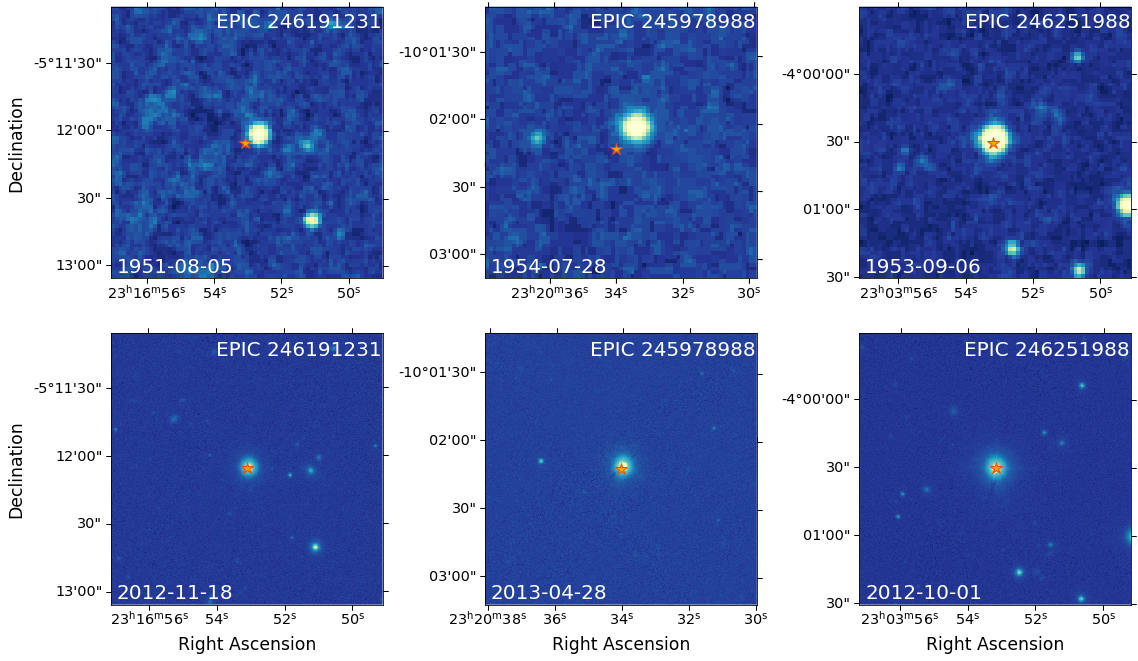} 
   \caption{A comparison of archival imaging observations of the three systems to determine whether any eclipsing binaries existed behind our three systems.  \textit{Top:} historic images of these systems from the 1950s taken with red-sensitive photographic plates during the Palomar Observatory Sky Survey conducted at Palomar Observatory in California. The orange stars in each image represent the present-day location of the targets. \textit{Bottom:} modern images taken with the Panoramic Survey Telescope and Rapid Response System (Pan-STARRS) with the i-band filter. We can see no stars in the present-day location of \staroneepic\ and \startwoepic, which eliminates the possibility that the transit signals are caused by a background eclipsing binary (see Section \ref{archivalimaging}). \starthree\ has too small of a proper motion to definitively claim the same .}
   \label{star_images}
\end{figure*}

\subsection{Spectroscopy} \label{spectroscopy}

We observed all three of our planet candidate hosting stars multiple times with the Tillinghast Reflector Echelle Spectrograph (TRES) on the 1.5 m telescope at Fred L. Whipple Observatory. TRES is a high resolution optical echelle spectrograph with a spectral resolving power of $\Delta \lambda / \lambda=44000$ and a limiting radial velocity precision of roughly 10 \ms. We used exposure times ranging from 700 seconds to 4500 seconds, with typical exposure times of 2000-3600 seconds.  Our team was able to acquire three spectra of \staroneepic, four spectra of \startwoepic, and five spectra of \starthree. These observations were spread over timescales of days to weeks, giving us enough baseline to identify stellar companions with periods similar to the likely orbital periods of our planet candidates. Because these sources are dim, the velocity precision is worse than the 10 \ms\ limit quoted above. We extracted radial velocity measurements for \starthree, a G-dwarf, using process outlined by \citet{gdwarfrv}, which involves cross-correlating each observed spectrum with the highest signal-to-noise observation using multiple spectral orders. We used used a similar method optimized for red stars (Pass et al. \textit{submitted}) to measure the velocities \staroneepic\ and \startwoepic, both of which are M-dwarfs. In brief, this method involves cross correlation of the TRES observation with a set of empirical templates using six echelle orders in the red-optical between 6400 and 7850 \AA. 
We use these radial velocity data to search for any evidence of a massive orbiting companions in Section \ref{radialvelocitymodelling}.  Our observations are summarized in Table \ref{rvvalues}.

\subsection{Speckle Imaging} \label{speckleimaging}

We obtained high-resolution imaging using the Zorro and \Alopeke\ speckle imagers at Gemini Observatory \citep{zorro}. Speckle images are created by processing thousands of exposures that are short enough to ``freeze'' the atmospheric turbulence that normally causes blurring and distortion of stellar images. These images are then processed individually and recombined to produce diffraction-limited images of small fields of view. This process is explained in depth in \citet{Speckle_Howell}. Speckle images allow us to see close stellar companions to targets of interest that may otherwise be difficult or impossible to detect. These short exposure images are taken using two color filters, 832 nm and 562 nm.

We observed \staroneepic\ and \starthree\ using the Zorro instrument at Gemini-South on October 23th, 2020 and November 30th, 2020 respectively. \startwoepic\ was observed by Gemini-North on June 28th, 2021. We processed these observations with the standard instrument pipelines \citep{Speckle_Howell}. 
For \staroneepic\ and \startwoepic, the speckle images yielded no detection of a stellar companion in either filter. For \starthree, however, the automatic speckle pipeline identified a possible detection of a faint red companion at a distance of 0.016 arcseconds from \starthree\ with a magnitude difference of 0.38. A visual inspection of the data revealed a weak, but plausible detection of a companion in the observation on November 30th, 2020. 

However, the detection of a possible companion to \starthree\ was made only by the 832 nm filter; no similar star was detected in the 562 nm filter. 
Because this possible companion was detected at Gemini Observatory's sensitivity limit and only detected in one filter, our team attempted to confirm this detection with a second speckle image of \starthree. In this second observation (taken on June 27th, 2021 with Gemini North), \starthree\ showed no evidence of a stellar companion. \bedit{Figure \ref{specklefigurestarthree} shows the contrast curves and images for the two observations.}

Because we only detected the possible companion to \starthree\ once and in a single imaging band, it proved difficult for our team to confirm or refute the detection with other observations. We first attempted to verify the detection by checking to see whether the \Gaia\ astrometric solution showed evidence for excess noise \citep[e.g.][]{evans, rizzuto2018, Belokurov}, which would reveal if \starthree\ moved as a star with a stellar companion would move. Unfortunately, given the small angular separation of the possible companion, \Gaia\ would not be able to detect a stellar companion at such a small separation \citep[see ][Figure 6]{Gaia_notes}. Another way to corroborate the existence of such a companion would be to search for the spectral lines of the companion in the TRES spectrum. The TRES spectrum of \starthree\ also showed no broadening or star blending, and therefore no evidence for the existence of the companion, but cannot rule it out either. Because the companion was detected only in the 832 nm filter, it would have to be fairly red, and could be significantly fainter than \starthree\ in the visible wavelengths where TRES is most sensitive. A final way to corroborate the existence of the companion would be to test whether the surface gravity of \starthree\ inferred from the TRES spectrum matches the surface gravity of the star we would infer from its absolute magnitude measured by \Gaia. If the companion to \starthree\ is real, we would expect it to appear brighter in \Gaia\ observations due to the extra brightness of the companion star, and therefore have a lower surface gravity than measured from the TRES spectrum. However, the gravity we measured from TRES was consistent with the luminosity measured by \Gaia, which indicates that the measured brightness of \starthree\ is consistent with that of a single star. This is suggestive that the companion may not exist, but given the large uncertainty on the magnitude difference between \starthree\ and the companion, we cannot rule out its existence for sure. 

Because we were unable to independently confirm or reject the existence of the companion, we consider different explanations for why we may have detected a companion once, but not a second time.  The first possibility is that the stellar companion has now moved in its orbit to be too close to \starthree\ to be resolved in our speckle images. It is also possible that the second image was simply not as sensitive (either in depth or inner working angle) as the original image that detected the companion. 
The final option, and the one we assume to be true for the remainder of this paper, is that the original detection was spurious. Because the detection was shown only in one wavelength image, very close to the telescope diffraction limit, does not appear in the second observation, and the brightness of \starthree\ is consistent with single star system with no companions, we concluded that the detection of a possible companion to \starthree\ was most likely spurious. For the rest of the paper, we treat \starthree\ in our analysis as a single star system. 

\subsection{TESS Light Curves}\label{tesslightcurves}

After detecting the transiting planet candidates around \staroneepic, \startwoepic, and \starthree, we proposed for Director's Discretionary Time (DDT) observations of the three planet candidates using NASA's \textit{Transiting Exoplanet Survey Satellite} (\TESS) mission (Program ID DDT044, PI: Incha). Like \Kepler, \TESS\ is a space telescope designed to make precise photometric observations of stars over a large area of sky. \TESS\ differs from \Kepler\ because it looks at a larger field of view than \Kepler, but has a smaller collecting area (\TESS\ uses 10 cm diameter cameras compared to \Kepler's 0.95 m diameter mirror), and therefore has lower photometric precision on the same stars compared to \Kepler. Moreover, \TESS's observing strategy is to observe one part of the sky for about 28 days at a time, before moving onto a different region, and eventually tiling the whole sky, so \TESS\ typically observes stars for less time than \Kepler\ \bedit{\citep{ricker}}. However, compared to the very short duration of observations from K2 Campaign 19, even a single set of \TESS\ observations would significantly increase the baseline of observations for these systems. 

\begin{figure}
      \centering
      \includegraphics[scale = 0.40]{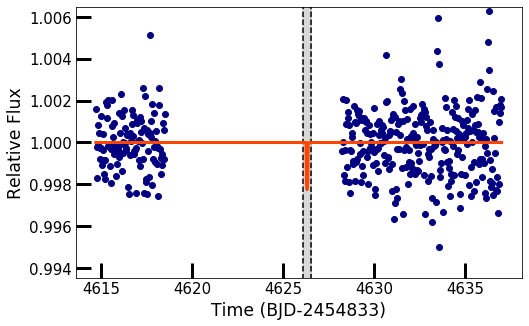}
      \caption{\TESS\ observations of \staroneepic. The \TESS\ light curve (binned to roughly 25 minute cadence for clarity) is shown as blue dots, and the best-fit transit model for \staroneepic\ b is shown in orange. The 1$\sigma$ uncertainty on the time of transit is shown as the grey region. Unfortunately, contamination from scattered light prevented \TESS\ from collecting useful data at the time we expected to see a transit based on the orbital period measured in Section \ref{multipletransits}.}
      \label{newtransitstessstarone}
   \end{figure}

\TESS\ observed the three stars during Sector 42 of its extended mission, and collected measurements of the stars' brightnesses every two minutes during the 28 day long sector. Unfortunately, this sector was adversely affected by scattered light entering the telescope from the Earth and Moon, and as a result, large portions of the dataset were unusuable. \TESS\ still managed to collect useful data for roughly 14 days for each planet, spread out over the course of 25 days. We used \TESS\ data processed by the \TESS\ Science Processing Operations Center (SPOC) pipeline, based at NASA Ames Research Center \citep{Jenkins2016}.  For \staroneepic\ and \starthree, we identified no evidence of transits in the \TESS\ data; the photometric precision was too low to detect the shallow transits of \starthree, and the only transit of \staroneepic\ predicted to happen based on the orbital period measured from the two transits found in K2 data happened to fall in a data gap caused by the scattered light (see Figure \ref{newtransitstessstarone}). \bedit{We attempted to recover light curves during the gaps caused by scattered light by analyzing the pixel-level data ourselves, (as done by \citealt{dalba2020}) using a custom FFI pipeline based on \citet{vanderburg19}, but were unable to recover useable data at those times.} 

\begin{figure*}
      \centering
      \includegraphics[scale = 0.50]{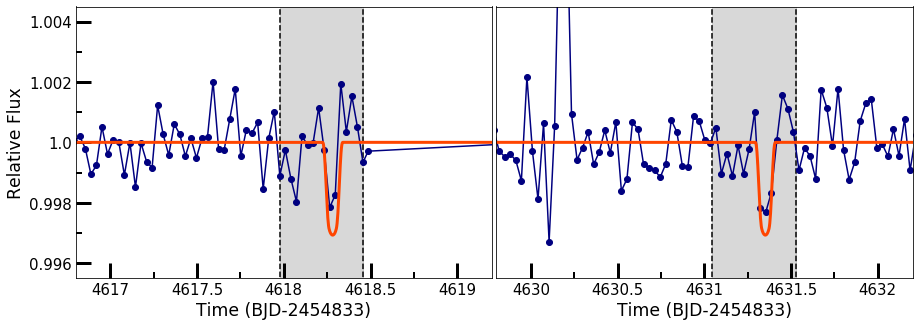}
      \caption{\TESS\ observations of \startwoepic. The \TESS\ light curve (binned to roughly 25 minute cadence for clarity) is shown as blue dots, the transit model for \startwoepic\ b is shown in orange (with the period adjusted slightly to match the timing candidate signal detected in TESS), and the 1$\sigma$ uncertainty of the predicted time of transits is shown as a grey region. We show the \TESS\ light curve zoomed near the two predicted transits of \startwoepic\ b during the observations. We detected two low-significance dips with depth, duration, and timing consistent with being transits of \startwoepic\ b, but wait for future observations to claim a definitive detection of the planet's transits. }
      
      \label{newtransitstessstartwo}
   \end{figure*}

We did see tentative evidence for two transits of \startwoepic\ in the \TESS\ data (see Figure \ref{newtransitstessstartwo}). These dips were separated by precisely the orbital period measured by K2, and happened within the 1 sigma uncertainty of the predicted time of transits based on the K2 ephemeris. The depth and duration of the TESS candidate transits are also consistent with the K2 transits, but with large uncertainties. Given the consistency of the timing and shape of the signals detected by TESS with those from K2, we consider this a tentative confirmation of the K2 detection, but do not include these data in our analysis for determining final parameters. 

\section{Analysis}\label{analysis}

\subsection{Stellar Parameters}\label{stellarparameters}

The first step of our analysis was to determine the fundamental stellar parameters of \staroneepic, \startwoepic, and \starthree. The initial characterization of the stars from the EPIC catalog revealed that \staroneepic\ and \startwoepic\ are likely M-dwarfs, and \starthree\ is likely a sun-like G-dwarf. We used different methods to characterize the two M-dwarfs and the G-dwarf.  For \starthree, a G-dwarf, we were able to derive spectroscopic parameters from the TRES spectra taken at Fred L. Whipple Observatory described in Section \ref{spectroscopy} using the Stellar Parameter Classification software \citep{SPCsoftware1, SPCsoftware2}. SPC yielded estimates of the star's effective temperature, surface gravity, and metallicity. From the spectroscopic parameters, the star's V-band magnitude from the TESS Input Catalog (TIC, \citealt{stassuntic}) and its \Gaia\ parallax \citep{gaiamission, gaiadr2}, we determined the stellar mass and radius using an online interface\footnote{\url{http://stev.oapd.inaf.it/cgi-bin/param_1.3}} that interpolates Padova stellar evolution models \citep{dasilva}. We calculated \starthree's luminosity using the Stefan-Boltzmann Law.

We took a different approach to determine stellar parameters for the two M-dwarf stars, \staroneepic\ and \startwoepic. Because it is more difficult to directly measure spectroscopic parameters for M-dwarfs from high-resolution optical spectra, we used empirical relationships described by \citet{Mann_2015} and \citet{Mann_2018} to determine each star’s fundamental parameters. In particular, we used the absolute K-band/mass relationship from \citet{Mann_2018}, the absolute K-band/radius relationship from \citet{Mann_2015}, and the V-K/temperature relationship from \citet{Mann_2015}. To calculate each star's parameters, our team used their K-band magnitudes, V-band magnitudes, and parallaxes as collected in the TIC. All of our results are listed in Table \ref{bigtable}. 

\subsection{Transit Modeling}\label{transitmodel}

We determined the best-fit planetary parameters and their uncertainties by fitting the \Kepler\ light curves with a transit model using Markov Chain Monte Carlo. We performed two separate analyses: 1) modeling the single transits of each planet candidate detected during the high-quality week of observations, and 2) an analysis for \staroneepic~b and \startwoepic~b also including the transits detected after \Kepler's fuel reserves began to expire. 

In both analyses, we modeled the shape of the transit light curves using the BATMAN transit modeling software \citep{batman}. By default, BATMAN calculates a model transit light curve based on nine parameters: the time of inferior conjunction, orbital period, planet/star radius ratio (\rprs), scaled semi-major axis ($a/r_\star$), orbital inclination, eccentricity, longitude of periastron, and two limb darkening coefficients. Often, it is most convenient to parameterize a transit model using orbital parameters like these, but here, we chose to parameterize our two analyses differently to speed convergence. \bedit{In both of our analyses, we also chose to supersample our data by a factor of 12 in order to account for the smearing of data that occurs due to \Kepler's 29.4 minute long exposure integration time.}

\subsubsection{Single Transits}\label{singletransitanalysis}

For our analysis of the single transits detected in the high-quality data, our strategy was to model the \textit{shape of the transit} in terms of the transit duration and impact parameter, rather than more degenerate orbital parameters like period, scaled semimajor axis, and inclination. Then, with the measured transit parameters in hand, we can use analytic relationships to calculate the corresponding distributions of orbital parameters. 

We constructed our model for single transits with the following parameters: the time of transit center, the natural log of the planet/star radius ratio, impact parameter, transit duration (from first to fourth contact), two limb darkening parameters, four parameters for a cubic polynomial describing the out-of-transit light curve variations, and the uncertainty on each \Kepler\ data point. We imposed priors on these quantities to ensure that the transit duration, planet radius, semi-major axis, impact parameter, first limb darkening parameter, and error are all greater than zero. We also imposed a prior that the natural log of the planet/star radius ratio must be less than zero to ensure that the planet radius is smaller than the star. We restricted the impact parameter to be less than $1+R_p/R_\star$ and we imposed informative Gaussian priors on the limb darkening parameters based on models computed by \citet{claretbloemen}. In particular, we imposed priors on $\{u1, u2\}$ of {$\{0.382\pm 0.152, 0.336 \pm 0.152\}$ for \staroneepic, $\{0.431\pm 0.152, 0.291 \pm 0.152\}$ for \startwoepic, and $\{0.253\pm 0.152, 0.289 \pm 0.152\}$ for \starthree.

Because we chose to parameterize our MCMC fit in terms of transit observables like duration and impact parameter, instead of the physical paramerization used by BATMAN, we had to convert our parameters to the ones used by BATMAN in our model. When considering a single transit that does not repeat, four of the physical parameters used by BATMAN (period, scaled semimajor axis, eccentricity, and argument of periastron) almost exclusively\footnote{In detail, orbital eccentricity can introduce a slight asymmetry in the shape of a transit, but this effect scales as $(a/R_\star)^{-3}$ and is negligible for the planets we consider here \citep{winn}.} affect the transit duration. Since we replaced these four parameters in our fit with a single parameter for the transit duration, we were able to fix and solve for the value of the fourth parameter that would yield a model transit with the desired duration.  We set eccentricity to 0, longitude of periastron to 90 degrees, and \bedit{because the period for this analysis is an arbitrary number that does not effect the final fit, we chose to set the period to  100 days, a nice round number large enough to guarantee that our code would not attempt to fit any out-of-transit data with a transit model.} With these three parameters fixed, we solved for the scaled semimajor axis, $a/r_\star$ following \citet{seager}:

\begin{equation}\label{semi_major_axis}
    \frac{a}{r_\star} = \sqrt{\frac{(1+(\rprs))^{2}-b^{2}[1-\sin^2{(\pi \frac{t_{14}}{p})}]}{\sin^2(\pi \frac{t_{14}}{p})}}
\end{equation}

\noindent where \rprs is the radius of the planet also in units of stellar radii, $b$ is the impact parameter, $t_{14}$ is the transit duration, and $p$ is the orbital period. We then solved for the orbital inclination from $a/r_\star$  and the impact parameter $b$: 

\begin{equation}\label{inclination}
    i = \arccos \left(\frac{b}{(a/r_{\star})} \right)
\end{equation}

Our MCMC was run using \texttt{edmcmc} \citep{edmcmc}, a Markov Chain Monte Carlo (MCMC) sampler that implements the Differential Evolution sampling algorithm described by \citet{terbraak}. In order to ensure convergence of the MCMC chains, we sampled parameter space using 40 chains and evolved each chain for 5,000,000 links. We tested convergence by calculating the Gelman-Rubin statistics for each parameter, all of which were below 1.01. We show the best-fit transit models for each planet candidate in Figure \ref{lightcurves_bestfits}, and our best-fit parameters and uncertainties are given in Table \ref{singletransitfitvalues}.

\subsubsection{Estimating Orbital Parameters from Single Transits}\label{period_calculation}

   
\begin{table*}
	\begin{center}
	\caption{Best-fit transit parameters and uncertainties from our single transit analysis.}
    \label{singletransitfitvalues}
	\begin{tabular}{cccc} 
		\hline
		Parameter & EPIC 246191231 & EPIC 245978988 & EPIC 246251988\\
		\hline
Orbital Period, $P$~[days] & \periodstarone & \periodstartwo & \periodstarthree \\
Radius Ratio, $r_P/r_\star$ & \rprstarone & \rprstartwo & \rprstarthree \\
Scaled semimajor axis, $a/r_\star$  & \arstarone & \arstartwo & \arstarthree \\
Semimajor axis [au], $a$  & \aaustarone & \aaustartwo & \aaustarthree \\
Orbital inclination, $i$~[deg] & \incstarone & \incstartwo & \incstarthree \\
Transit impact parameter, $b$ & \impactstarone & \impactstartwo & \impactstarthree \\
Transit Duration, $t_{14}$~[hours] & \transitdurationstarone & \transitdurationstartwo & \transitdurationstarthree \\
Time of Transit, $t_{t}$~[BJD] & \transittimestarone & \transittimestartwo & \transittimestarthree \\ 
Planet Radius, $R_P$~[\rearth] & \rpstarone & \rpstartwo & \rpstarthree \\
Equilibrium Temperature, $T_{eq}$~[K] & $503.2 \pm 9.2$ & $548.6 \pm 6.1$ & $868.0 \pm 15.3$ \\
		\hline
	\end{tabular}
	\end{center}
	
	\textit{Notes: } The values reported here for \staroneepic\ and \startwoepic\ are superceded by the fits reported in Table \ref{bigtable}, which include additional transits to precisely measure the orbital period.
\end{table*}

After determining the best-fit transit parameters and uncertainties with our MCMC analysis, we turned our attention to the task of extracting orbital parameters from the results. Our MCMC yielded well-converged posterior distributions of the transit duration, star/planet radius ratio, and impact parameter. We started by calculating the orbital period inferred from each posterior draw following \citet{zeit}:

\begin{equation}\label{period}
     p = \left(\frac{t_{14}(\frac{\pi}{4} G M_{\star})^{1/3}}{\sqrt{\left \lvert(r_{p}+r_{\star})^{2}-(b^{2} r_{\star}^{2})\right \rvert}}\times\frac{1+(e\cos(\omega))}{\sqrt{\lvert1-e^{2}\rvert}}\right)^{3}
\end{equation}

\noindent where $p$ represents the orbital period of the planet, $G$ is the gravitational constant, $M_{\star}$ is the mass of the star, $r_{p}$ is the radius of the planet, $r_{\star}$ is the radius of the star, and $b$ is the impact parameter. We generated random samples of the stellar mass and radius based on the results of our analysis in Section \ref{stellarparameters}. To account for the unknown eccentricity $e$ and argument of periastron $\omega$, we randomly generated values using the \texttt{ECCSAMPLES} software \citep{eccsamples}. 

Then, we adjusted our orbital period distributions to remove orbital periods ruled out by the observations and de-emphasized orbital periods unlikely to be detected by our observations. First, we used the fact that we detected only a single transit of each planet in the 7.25 days of observations to impose a hard lower-bound on the possible orbital periods. We identified the longest time period during which K2 observed before or after each transit. For example, the period of the planet candidate orbiting \staroneepic, as shown in Figure \ref{lightcurves_bestfits}, must be greater than about 5 days, since K2 observed no transits in the time interval after the transit day 3 to day 8. Once we identified the longest period of time either before or after the transit when K2 was observing for each planet candidate, we removed orbital periods shorter than this interval from our posterior.

We also accounted for the fact that the longer the planet candidate's orbital period, the less likely it is that we would have detected a transit during the relatively short duration of the K2 observations (in particular, the week of high-quality data we used to detect the candidates). The probability $\mathcal{P}$ of detecting at least one transit during an observational baseline $B$ is given by: 

\begin{equation}\label{priordetection}
\mathcal{P} = \frac{(B + t_{14})}{p} \text{  for  } p > (B +  t_{14})
\end{equation}

\noindent where $p$ is the orbital period and $t_{14}$ is the transit duration. We randomly discarded samples with orbital periods longer than the observational baseline plus the transit duration for each planet candidate by calculating a random number uniformly distributed between 0 and 1, and removing it if it was larger than $\mathcal{P}$ for that orbital period. 

The final posterior probability distributions for the orbital period of all three candidates (based on the single transits observed during the week of high quality data) are shown in Figure \ref{habitablezones}. Based on this analysis, all three planet candidates likely have relatively short orbital periods, with the probability distributions peaking at periods slightly longer than our observational baseline, so it is not surprising that inspection of the lower-quality data would have yielded additional transits in some of these cases. The measured orbital periods for the two candidates with multiple observed transits (\staroneepic~b and \startwoepic~b) are consistent with the posterior distributions derived here. With our final posterior probability distributions for orbital period in hand, we used Equations \ref{semi_major_axis} and \ref{inclination} to calculate the corresponding distributions of semimajor axes and inclinations.

\begin{figure*}
      \centering
      \includegraphics[scale = 0.5]{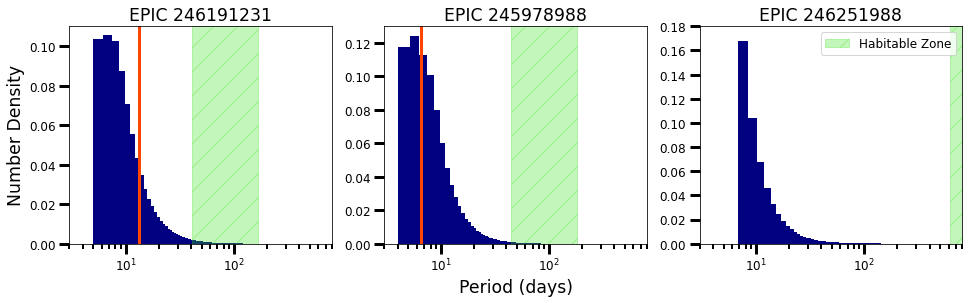}
      \caption{Posterior probability distributions for the orbital period of each planet candidate, based on our analysis of the single transits in Section \ref{period_calculation}. We overlaid the location of the optimistic habitable zones of each star (as described by \citealt{habitable1}). We also show orange vertical lines at the true orbital periods of \starone\ b and \startwo\ b. The measured orbital periods are consistent with our estimates from the single transit analysis.}
      \label{habitablezones}
   \end{figure*}

\subsubsection{Multiple Transits}\label{multipletransits}

We detected additional transits for two of our planet candidates, \staroneepic~b and \startwoepic~b, and therefore were able to analyze the two transits together to precisely measure the orbital period and sharpen our knowledge of other parameters thanks to the additional in-transit data. We made slightly different choices in the parameterization and assumptions in this analysis compared to the single-transit analysis described in Section \ref{singletransitanalysis}.

We parameterized our model as follows: period, time of mid-transit, impact parameter, log of the ratio between the planet and star radius ($\log{(R_p/R_\star)}$), transit duration, linear and quadratic limb darkening parameters, and a flux offset parameter. To cut down on the number of free parameters and speed model convergence, we did not fit for normalization/out-of-transit variability simultaneously with the transit parameters in our MCMC model, as we had previously done in our single-transit analysis. Instead, we subtracted the best-fit low-frequency variability signal from our simultaneous transit/systematics/variability fit described in Section \ref{lightcurve}. We note that even though we tentatively detected the transits of \startwoepic~b in \TESS\ data, we do not include those data in this fit, given the relatively low-signal-to-noise ratio of the detections. We imposed the same priors on limb darkening in these fits as we did in the ones described in Section \ref{singletransitanalysis}. 

We ran the fit using \texttt{edmcmc}, evolving 40 independent chains over 1,000,000 links using the differential evolution algorithm of \citet{terbraak}. We measured Gelman-Rubin values less than 1.001 for all parameters, indicating that the chains were well converged. With these well-converged chains in hand, we used similar procedures to those described in Section \ref{period_calculation} to derive posterior distributions of astrophysical parameters like $a/R_\star$, although much of that analysis was no longer needed since we directly fitted for the orbital period. We report the best-fit values and uncertainties from this updated analysis for \staroneepic\ and \startwoepic\ in Table \ref{bigtable}.

\subsection{Radial Velocity Modelling} \label{radialvelocitymodelling}

Shallow transits can sometimes be mimicked by grazing eclipses of orbiting binary stars. To rule out this scenario as the origin of the transit signals in the three systems we identified, we analyzed radial velocity observations to place limits on the masses orbiting bodies that might produce the transits we see. 

We fitted the radial velocities measurements with Keplerian orbit models (as implemented by the \texttt{Radvel} package \citealt{radvel}) using MCMC. We performed two different fits: one assuming circular orbits, and one allowing eccentric orbits. For circular orbits, we had only four free parameters: the period, time of mid-transit, the radial velocity semiamplitude, and the velocity zero-point. For eccentric orbits, we added combinations of eccentricity, $e$, and the argument of periastron, $\omega$ as two additional free parameters: $\sqrt{e}\cos{\omega}$ and $\sqrt{e}\sin{\omega}$. We imposed priors on the period and time of transit based on the results of our transit analyses. For \staroneepic\ and \startwoepic, we imposed Gaussian priors with center-points and widths based on the results of our multi-transit fits, while for \starthree, we sampled directly from the orbital period posterior probability distribution from Section \ref{period_calculation}. 

We performed the fits using \texttt{edmcmc} \citep{edmcmc}, evolving 20 independent chains for 1 million links. We assessed convergence using the Gelman-Rubin statistic \citep{gelmanrubin}, which had values below 1.01 for each parameter. We show the radial velocities along with our-best-fit orbits and randomly selected orbits drawn from the posterior distribution in Figure \bedit{\ref{thesinegraphs}, which includes figures for both circular and eccentric fits.} We converted the posterior distributions in semiamplitude, period, and eccentricity into a posterior distribution in mass and placed limits on the mass of orbiting companions. We summarize our mass limits in Table \ref{mpmjvalues}. 

We were able to rule out the possibility of an orbiting star in all cases with 3$\sigma$ confidence, and were also able to rule out an orbiting brown dwarf to a three sigma certainty for \staroneepic\ and \startwoepic. Our constraints for \starthree\ are weakest because it lacks a precisely measured orbital period, and we struggle to confidently rule out even high-mass brown dwarfs in the long-period tail of the probability distribution. Nevertheless the preponderence of the evidence points to a planetary mass for any companion orbiting \starthree.

\subsection{False Positive Probability Calculation/Validation}\label{validation}

Finally, we calculated the astrophysical false positive probability of two of our planet candidates: \staroneepic\ b and \startwoepic\ b. This type of analysis is often employed to quantify the likelihood that any given candidate is a real exoplanet. Typically, if the astrophysical false positive probability is found to be below some threshold, and all other diagnostics (such as those described in previous sections) show no evidence that the signals are false positives, a candidate can be considered ``statistically validated'' and treated as a highly likely planet. 

We chose to focus on these candidates (and not \starthree\ b) because they both have two observed transits and therefore precisely measured orbital periods. Seeing multiple transits is highly important for statistical validation for two reasons. First, observing at least two high signal-to-noise transits that are consistent in shape greatly decreases the likelihood that they are instrumental artifacts. A single data anomaly could plausibly mimic a convincing single transit, but it is much less likely that this could happen twice in the light curve, without also producing other artifacts that could be identified. Second, knowledge of the true orbital period of the planet candidates provides an important constraint on false positive scenarios by confirming that the duration and shape of the transit is reasonable given the orbital period and known properties of the presumed stellar host. Cases where the transit duration and shape are very different from what might be expected for a planet orbiting the host star are more likely to be false positives. 

\begin{figure*}
      \centering
      \includegraphics[scale = 0.75]{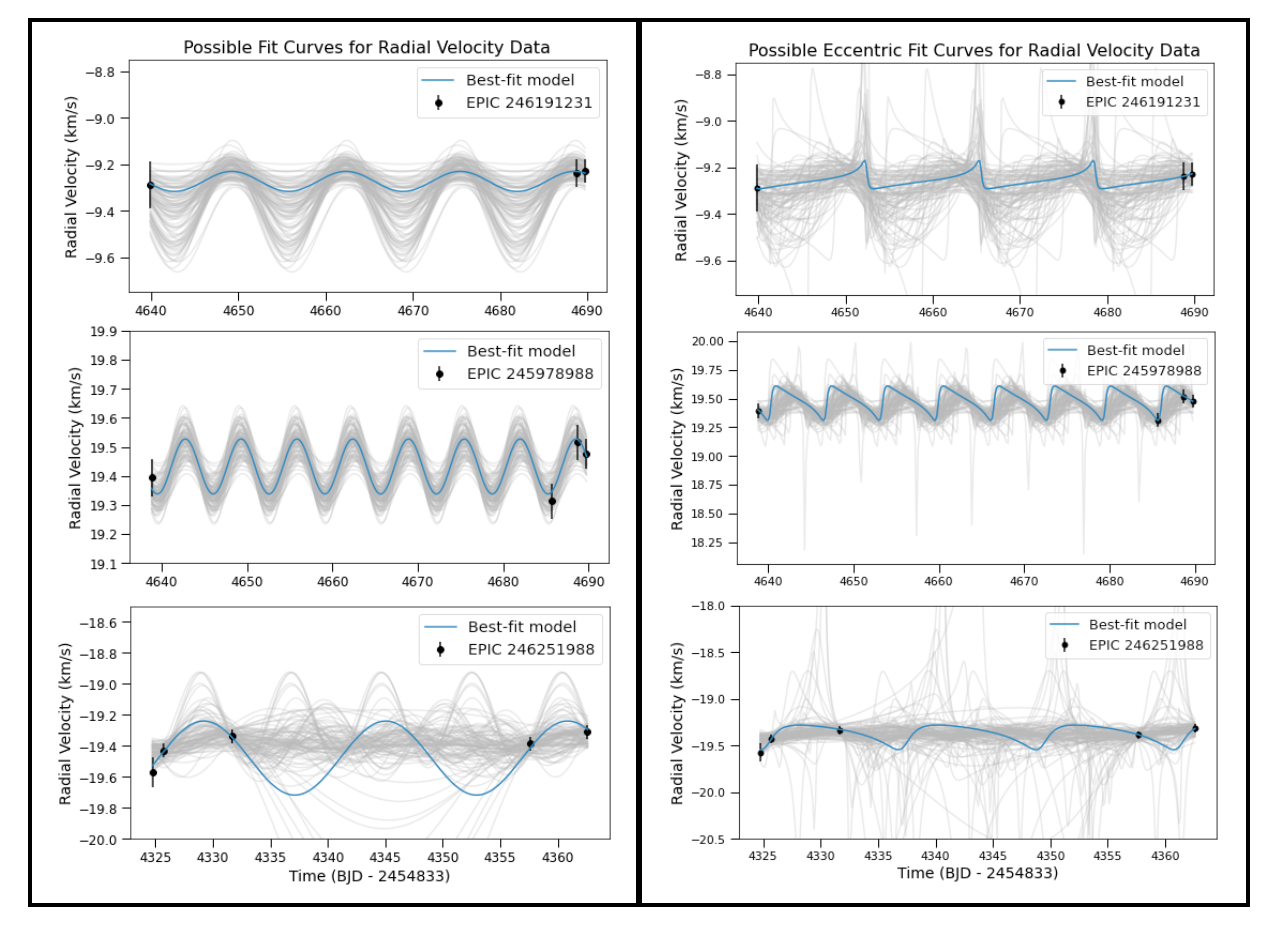}
      \caption{ \bedit{Left: Radial velocity observations of the three stars compared to models drawn from our MCMC analysis allowing only circular orbits. The individual TRES radial velocity observations are shown as black points with error bars. In grey, we plot models corresponding to 100 random draws from our MCMC posterior, and in blue, we plot the best-fit RV model. The RV observations can confidently rule out stellar companions in all cases, and rule out brown dwarf masses with >3$\sigma$ confidence for \staroneepic\ b and \startwoepic\ b. Right: Radial velocity observations of the three stars compared to models drawn from our MCMC analysis allowing non-zero eccentricity. Like in the figure at left, the TRES radial velocity measurements are shown as black points, models from randomly selected posterior draws are shown as grey curves, and the best-fit model is a blue curve. Even when the assumption of zero eccentricity is relaxed, we confidently rule out stellar and brown dwarf mass companions for \staroneepic\ and \startwoepic, and still rule out stellar-mass companions for \starthree, though the mass upper limits are somewhat weaker.}}
      \label{thesinegraphs}
   \end{figure*}

\begin{figure*}
      \centering
      \includegraphics[scale = 0.45]{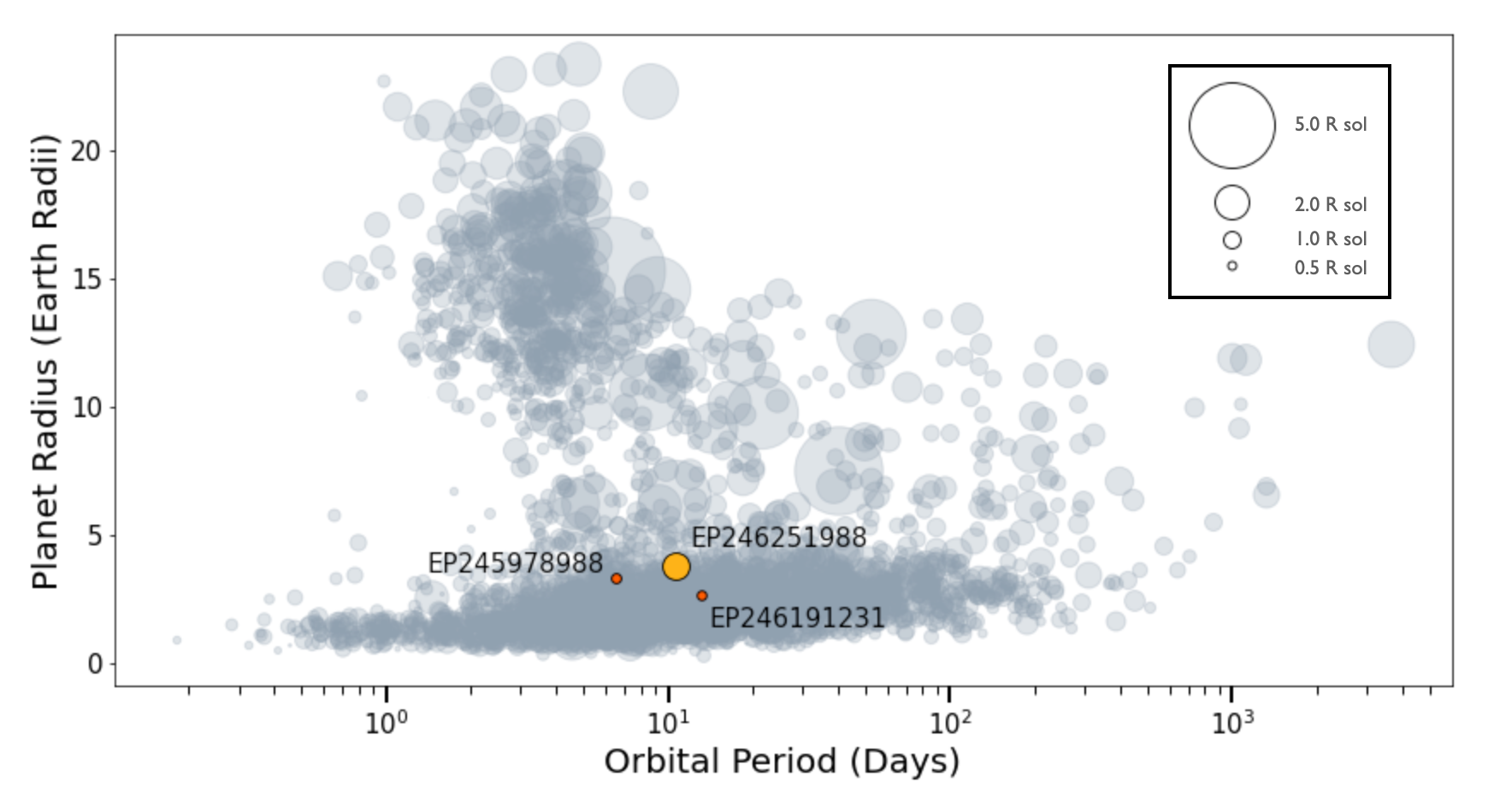}
      \caption{The population of known transiting exoplanets compared to \starone\ b, \startwo\ b, and \starthree\ b. We show the population of exoplanets in period/radius space, since these are the most readily measured observables for transiting planets like these objects. The grey points are planets other than the ones shown here, and the size of the symbol is proportional to the size of the planet's host star. \starone\ b and \startwo\ b are shown in red, since their hosts are M-dwarfs, while \starthree\ b is shown in orange to represent its G-dwarf host. The periods and radii of these three planets are typical of other small transiting planets discovered by \Kepler\ and \TESS. }
      \label{radiusperiodplot}
   \end{figure*}

We used the \texttt{vespa} software tool \citep{morton2015} to calculate the false positive probability of \staroneepic\ b and \startwoepic\ b. Given inputs such as the transit light curve, spectroscopic parameters for the host star, and constraints from high-resolution imaging observations, \texttt{vespa} performs Bayesian model comparison to calculate the likelihood that a planet candidate signal is indeed caused by an orbiting exoplanet compared to the likelihood of a variety of false positive scenarios. \texttt{Vespa} is an implementation of the false positive calculation described by \citet{morton12} and \citep{morton16}. As inputs, we fed the K2 light curves for \staroneepic\ and \startwoepic, the candidate orbital periods and radii, contrast limits from our speckle observations, broadband photometry from the 2MASS survey and the \Gaia\ parallax, and the star's location in the sky.

\texttt{Vespa} calculated very low astrophysical false positive probabilities for both stars (of roughly 10$^{-4}$ for \staroneepic\ b and an implausibly low value of 10$^{-10}$ for \startwoepic). These low probabilities are only strengthened by our additional constraints from radial velocities that rule out radial velocity variations from eclipsing binary companions and any background stars at the present-day location of \staroneepic\ and \startwoepic. We therefore conclude that there are vanishingly few plausible astrophysical false positive scenarios that can explain the transit signals seen at \staroneepic\ and \startwoepic. Given that we found no indication that the transit signals were instrumental artifacts, that each K2 light curve shows two transits with consistent shape, and that the \TESS\ light curve of \startwo\ shows evidence of two additional transits, we conclude that \staroneepic\ b and \startwoepic\ b are both overwhelmingly likely to be real planets consider them to be statistically validated,\bedit{ and hereafter refer to them as \starone\ b and \startwo\ b.}

\section{Discussion}\label{discussion}


In this work, we have discovered three planet candidates in the last observations from the \Kepler\ space telescope, and shown that two of them are very likely to be real planets through statistical validation. Despite the unusual circumstances of their discovery, these planets are all fairly typical of the wider population of sub-Neptune-sized objects discovered by the transiting method, as shown in Figure \ref{radiusperiodplot}. The three planets all have radii greater than 2.5 \rearth, putting them in the population of planets likely to host at least some volatile envelope \citep[e.g.][]{notrocky, gapinradii}. We estimated the masses of the three planets using the relationships derived by \citet{massfromradius} to be $7.3\ensuremath{^{+6.1}_{-3.2}}$ \mearth\ for \starone\ b, $9.5\ensuremath{^{+7.0}_{-3.4}}$ \mearth\ for \startwo\ b, and $13.9\ensuremath{^{+13.1}_{-7.3}}$ \mearth\ for \starthree\ b.

Given the short baseline of observations we used to detect the three single-transit candidates, it is unsurprising that they all very likely have short orbital periods and very high equilibrium temperatures. In two of the three cases (\starone\ and \startwo), we were able to confirm periods of less than 15 days using data collected after \Kepler\ began to run out of fuel, and while we did not detect an additional transit of \starthree\ b in these data, the probability distribution peaks at orbital periods just slightly longer than the duration of K2 Campaign 19 observations.

As such, all three planets/candidates are either confirmed to be, or very likely to be much hotter than the Earth and neither they (nor any moons they may host) are likely to be hospitable places for life as we know it to exist. We estimated the equilibrium temperatures, $T_{eq}$ of the three planets using the following expression \footnote{We note that the equilibrium temperatures are approximate because although this expression assumes perfect heat redistribution, these planets are likely tidally locked due to their short orbital periods.} 

\begin{equation}\label{massradiusrelation}
     T_{eq} = T_{\star} \sqrt{\frac{R_{\star}}{2a}}(1-A_{B})^{1/4}
\end{equation}

\noindent where $T_{\star}$ is the stellar temperatures, $R_{\star}$ is the stellar radius, $a$ is each planet candidate's semi-major axis  and $A_{B}$ is the planet candidate's albedo, which we estimate to be $0.05 \pm 0.04$ using the values found by \citet{albedo}. We found equilibrium temperatures of $503.2 \pm 9.2 K$, $548.6 \pm 6.1 K$, and $868.0 \pm 15.3 K$, for \starone\ b, \startwo\ b, and \starthree\ b, respectively. 

\begin{table}
	\begin{center}
	\caption{Upper limits on the masses of our candidate companions.}
\label{mpmjvalues}
	\begin{tabular}{lcccc} 
		\hline
		Star & 2$\sigma$ & $2\sigma$ & $3\sigma$ & $3\sigma$\\
		(EPIC ID) & (circular) & (eccentric) & (circular) & (eccentric)\\

		\hline
246191231 & 1.7 \mj & 2.4 \mj & 2.5 \mj & 5.2 \mj \\
245978988 & 1.0 \mj & 1.3 \mj & 1.3 \mj & 2.2 \mj \\
246251988 & 4.3 \mj & 12.3 \mj & 17.0 \mj & 27.9 \mj\\
		\hline
	\end{tabular}
	\end{center}
	
\end{table}

We then calculated the location of the habitable zone using the calculations of \citet{habitable1} for the three stars and compared them to the orbits we derived in the three systems. We show the results of these calculations in Figure \ref{habitablezones}. While the long-period tails of the original posterior probability distributions for the orbits of \starone\ b and \startwo\ b overlapped with the habitable zones in those systems, the detection of additional transits confirmed orbital periods well interior to the habitable zone. Meanwhile, even though we have not yet measured the orbital period of \starthree\ b, even the long-period tail of probability distribution for the planet candidate's orbit is well interior to the habitable zone.

Unlike the vast majority of the planets found by \Kepler, the three planet/candidates detected in this work were all found originally as single transit discoveries. This was necessitated because of the very short duration of observations taken during K2 Campaign 19; with only 7.25 days of high-quality data, only planets with periods shorter than about 3.6 days are guaranteed to transit at least twice. According to the NASA Exoplanet Archive, about 81\% of all exoplanets known today have orbital periods longer than this value, so the vast majority of planets to be found in the K2 Campaign 19 dataset were likely to have only transited once. 

Single-transits are a powerful way to detect planets with longer orbital periods than the duration of observations searched, but they have significant downsides. It is more challenging to detect them than it is to detect multiply transiting planets, and often involves highly-tuned pipelines \citep[e.g.][]{foremanmackey} or careful visual inspection \citep[e.g.][]{kristiansen}, as done in this work. The resulting planetary parameters, especially the orbital period and semimajor axis, are much less precise, and detecting additional transits is challenging, even with many follow-up observations \citep[e.g.][]{singletransittess1}. 


Despite these challenges, single-transits are likely to grow in importance. Currently, the \TESS\ mission is observing stars for only about a month at a time, so the majority of its discoveries at long periods will be detected at first as a single transit. Methods like the ones used in this paper can be valuable tools going forward to analyze and prioritize temperate planet discoveries from TESS. 


\begin{table*}
	\begin{center}
	\caption{System parameters for our two planets and one planet candidates. }
\label{bigtable}
	\begin{tabular}{lccc} 
		\hline
		Parameter & \starone\ & \startwo\ & \starthree \\
		 & & & \\
		\hline
\emph{Designations}  & \\
EPIC ID  & 246191231 & 245978988 & 246251988 \\
TIC ID   & 301132473 & 49592509 & 248050219 \\
\\
\emph{Basic Information} & \\
Right Ascension & 23:16:52.989  & 23:20:33.884 & 23:03:53.172 \\
Declination & -05:12:04.54  & -10:02:09.49 & -04:00:30.06 \\
Proper Motion in RA [\ensuremath{\rm mas\,yr^{-1}}]& $91.366 \pm 0.068$ & $133.632 \pm 0.063$ & $9.61542 \pm 0.074$  \\
Proper Motion in Dec [\ensuremath{\rm mas\,yr^{-1}}]& $-56.165 \pm 0.052$ & $-144.15 \pm 0.045$ & $-18.8167 \pm 0.053$  \\
Distance to Star~[pc] & \distancestarone & \distancestartwo & \distancestarthree \\
\Gaia\ G-magnitude & 13.9468 $\pm$ 0.00053  &  13.1533 $\pm$  0.00059 & 11.965 $\pm$  0.00023 \\ 
K-magnitude & $10.734 \pm 0.021$ & $10.06 \pm 0.019$ & $10.494 \pm 0.019$ \\ 
\\
\emph{Stellar Parameters} & \\
Mass, $M_\star$~[$M_\odot$] & \massstarone & \massstartwo & \massstarthree  \\
Radius, $r_\star$~[$R_\odot$] & \radiusstarone & \radiusstartwo & \radiusstarthree  \\
Surface Gravity, $\log g_\star$~[cgs] & \loggstarone & \loggstartwo & \loggstarthree  \\
Effective Temperature, $T_{\rm eff}$ [K] & \efftempstarone & \efftempstartwo & \efftempstarthree \\
Luminosity [$L_\odot$] & \luminositystarone & \luminositystartwo & \luminositystarthree\\
 & & \\
Orbital Period, $P$~[days] & \periodstaronesecondtransit & \periodstartwosecondtransit & \periodstarthree  \\
Radius Ratio, $r_P/r_\star$ & \rprstaronesecondtransit & \rprstartwosecondtransit & \rprstarthree  \\
Scaled semimajor axis, $a/r_\star$  & \arstaronesecondtransit & \arstartwosecondtransit & \arstarthree  \\
Semimajor axis [au], $a$  & \aaustaronesecondtransit & \aaustartwosecondtransit & \aaustarthree  \\
Orbital inclination, $i$~[deg] & \incstaronesecondtransit & \incstartwosecondtransit & \incstarthree  \\
Transit impact parameter, $b$ & \impactstaronesecondtransit & \impactstartwosecondtransit & \impactstarthree  \\
Transit Duration, $t_{14}$~[hours] & \transitdurationstaronesecondtransit & \transitdurationstartwosecondtransit & \transitdurationstarthree  \\
Time of Transit, $t_{t}$~[BJD] & \transittimestaronesecondtransit & \transittimestartwosecondtransit & \transittimestarthree \\ 
Planet Radius, $R_P$~[\rearth] & \rpstaronesecondtransit & \rpstartwosecondtransit & \rpstarthree  \\
Equilibrium Temperature, $T_{eq}$~[K] & $447 \pm 49$ & $558 \pm 55$ & $866 \pm 263$ \\
		\hline
	\end{tabular}
	\end{center}
\end{table*}

The fact that these three planets/candidates were originally detected as single transits is not the only reason it may be difficult to learn more about them. Another important limitation to future studies is that like most planets found by \Kepler, none of the three planets/candidates in this work orbit particularly bright host stars. The \Gaia\ G-band magnitudes of the three planet/candidate hosts range from around 12 to 14, which makes them challenging targets for follow-up observations like precise radial velocities that have been used to measure the orbital periods of single-transiting planets in the past \citep[e.g.][]{HIP116454, santerne19, dalba}. The faintness of the hosts will also make it challenging for the TESS mission to detect additional transits to confirm or measure the planets' orbital periods. The photometric precision achieved by TESS is likely sufficient to detect individual transits of \starone\ b and \startwo\ b (at low significance), but likely too poor to detect individual transits of \starthree, the one candidate remaining without a precisely measured orbital period. It may take many additional observing sectors from future TESS extended missions to detect the shallow transits of \starthree\ b\bedit{, although currently there are no additional observations of these stars scheduled with TESS through the end of the fifth year of observations.}

While the planets/candidates reported in this work are not likely to be studied extensively in the future, they still are important historically because of the circumstances of their discovery. These are the only planets that have been discovered so far from K2 Campaign 19, and therefore strong contenders for the last planets ever detected chronologically over the course of the mission. Given the tremendous influence of \Kepler\ on the field of exoplanets, and more broadly, on humanity's understanding of our place in the cosmos, searching for more planets from \Kepler's final dataset is of interest to both fully exploit the telescope's data and learn the identity \bedit{of} its last planet discoveries. 






\section{Summary}\label{summary}


The \Kepler\ space telescope was responsible for the discovery of over 2,700 confirmed exoplanets during both its primary and extended K2 mission. In this paper we provide a bookend to these discoveries by describing three planet candidates discovered during K2's final observation set, Campaign 19. Due to its short duration, each of these planet candidates had only one recorded transit during the 7.25 days of high-quality Campaign 19 observations, though two of them showed an additional transit in lower-quality data taken after fuel reserves began to run out.

Two of the three planet candidates were discovered around \starone\ and \startwo, M-dwarfs of masses \massstarone\ $M_{\odot}$ and \massstartwo\ $M_{\odot}$ respectively and radii \radiusstarone\ $R_{\odot}$ and \radiusstartwo\ $R_{\odot}$ respectively. The third was discovered around \starthree, a sun-like G-dwarf of mass \massstarthree\ $M_{\odot}$ and radius \radiusstarthree\ $R_{\odot}$. We determined the planet candidates' most likely planetary parameters, such as their radii (\rpstarone\ \rearth, \rpstartwo\ \rearth, and \rpstarthree\ \rearth\ respectively) and their semimajor axes (\aaustarone\ au, \aaustartwo\ au, and \aaustarthree\ au respectively) using MCMC modeling. These planet candidates are likely hot with extended gaseous envelopes and uninhabitable.

Even though we only were able to observe a single transit of one of the planet candidates, we were able to use modern modelling techniques to determine most of its planetary parameters with useful precision.  This is analogous to the work being done with single-transit planet candidates identified in TESS data, since TESS also observes much of the sky for relatively short intervals and detects a large number of single transit candidates \citep[e.g.][]{singletransittess1, singletransittess2}. Our work runs in parallel to work being done by the TESS single-transit working group to determine these planetary characteristics from similarly small amounts of data. 

Like most of the planets detected by \Kepler, these three candidates orbit faint host stars, which makes detailed follow-up observations, like radial velocity mass measurements, challenging. It may therefore be difficult to learn much more about these planetary systems. In the case of \starthree, however, there is a chance that future \TESS\ observations may be able to recover the planet candidate's precise orbital period and allow a statistical validation of the system. 

These three systems represent the first planet candidates found in K2's Campaign 19 and some of the final planet candidates ever to be discovered by the \Kepler\ Telescope. As some of the last discoveries ever made by the \Kepler\ space telescope, these planet candidates hold significance both sociologically and scientifically. We hope our work will help ensure that all \Kepler\ data is utilized to its full potential so that no planets are left behind.

\section*{Acknowledgements}

This research has made use of NASA's Astrophysics Data System, the NASA Exoplanet Archive, which is operated by the California Institute of Technology, under contract with the National Aeronautics and Space Administration under the Exoplanet Exploration Program, and the SIMBAD database,
operated at CDS, Strasbourg, France. 

This paper includes data collected by the TESS mission. Funding for the TESS mission is provided by the NASA's Science Mission Directorate. This publication makes use of data products from the Two Micron All Sky Survey, which is a joint project of the University of Massachusetts and the Infrared Processing and Analysis Center/California Institute of Technology, funded by the National Aeronautics and Space Administration and the National Science Foundation. This paper includes data collected by the \Kepler\ mission. Funding for the \Kepler\ mission is provided by the NASA Science Mission directorate. Some of the data presented in this paper were obtained from the Mikulski Archive for Space Telescopes (MAST). STScI is operated by the Association of Universities for Research in Astronomy, Inc., under NASA contract NAS5--26555. Support for MAST for non-HST data is provided by the NASA Office of Space Science via grant NNX13AC07G and by other grants and contracts. This work has made use of data from the European Space Agency (ESA) mission {\it Gaia} (\url{https://www.cosmos.esa.int/gaia}), processed by the {\it Gaia} Data Processing and Analysis Consortium (DPAC, \url{https://www.cosmos.esa.int/web/gaia/dpac/consortium}). Funding for the DPAC
has been provided by national institutions, in particular the institutions participating in the {\it Gaia} Multilateral Agreement.\\

\section*{Data Availability}

We will upload all datasets used to Exofop upon acceptance of the manuscript.











\bsp	
\label{lastpage}
\end{document}